# *In vivo* optogenetic identification and manipulation of GABAergic interneuron subtypes

Lisa Roux[1], Eran Stark[1], Lucas Sjulson[1,2] and György Buzsáki[1]

1. NYU Neuroscience Institute, School of Medicine and Center for Neural Science, New York University, New York, NY 10016

2. Department of Psychiatry, New York University Medical Center, New York, NY 10016

E-mail: gyorgy.buzsaki@nyumc.org

## Abstract

Identification and manipulation of different GABAergic interneuron classes in the behaving animal are important to understand their role in circuit dynamics and behavior. The combination of optogenetics and large-scale neuronal recordings allows specific interneuron populations to be identified and perturbed for circuit analysis in intact animals. A critical aspect of this approach is the spatial and temporal precision required for light delivery coupled with electrophysiological recording. Focal multisite illumination of neuronal activators and silencers in predetermined temporal configurations or a closed loop manner opens the door to addressing many novel questions. Recent progress demonstrates the utility and power of this novel technique for interneuron research.

**Introduction**

Computation in neural networks relies on dynamic interactions between excitatory and inhibitory cell types [1-4]. Appropriately timed inhibition targeted to specific somatodendritic domains of principal cells selectively filters synaptic excitation and modulates the gain, timing, tuning and bursting properties of pyramidal cell firing [4, 5]. Inhibitory interneurons also secure the transient autonomy of principal cells by flexibly congregating and segregating neuronal populations (often referred to as cell assemblies) through maintenance of oscillations and synchrony [6, 7]. A large variety of inhibitory interneurons are available for such tasks [2, 3, 5, 8, 9], but the specific contributions of the different subtypes are still unclear. Our understanding of neural network functions could therefore be advanced greatly by studying the activity of specific interneuron subtypes in intact circuits and by perturbing them in a targeted manner [10-17].

Several methods have been proposed to identify specific interneuron classes in extracellular recordings, including physiological classification based on spike waveforms, firing patterns and network-affiliated activity [7, 18-21]. Alternatively, intracellular or juxtacellular recordings, either blind [2, 9, 22, 23] or fluorescence-targeted [12, 24-27], have provided valuable information about the roles of various interneuron classes in the generation of network activity [2, 9]. Here we discuss how optogenetic approaches, combined with large-scale extracellular recordings in behaving animals, can be used to identify and manipulate interneurons to facilitate the understanding of their computational roles in neural circuits.

**Optogenetic identification of interneuron subtypes**

Optogenetics [28-30] provides a solution for identifying specific neuronal subtypes in blind extracellular recordings by expressing light-sensitive opsins in a given neuronal population and inferring that light-responsive units correspond to members of that population. Both activation [31-33] and silencing [34] strategies can be used for this purpose. Various methods for cell-type specific expression of opsins have been developed [35-38] along with techniques to deliver optical stimulation during extracellular electrophysiological recordings [39-42]. While implementing this photostimulation-recording method seems relatively straightforward, a number of technical issues must be addressed to exploit its full potential.

*Drawbacks of current photostimulation methods*

Commercially available optical fibers used in most current optogenetic experiments typically have large diameters (>100 µm) and should be placed at least 200 µm away from the recording electrodes to avoid damage of the targeted neuronal population [43]. Due to the strong light-absorbing nature of neuronal tissue, high light power (several mW) is therefore required to activate neurons at the distant recording sites [43-45, 46 ]. Brain surface illumination requires even higher power (>30 mW) to reach deeper cortical layers [47, 48]. Continuous application of light at such high power can cause local heating, leading to neuronal dysfunction and potential cellular damage.

Moreover, high power photostimulation can prevent accurate "spike sorting." This key process in large-scale extracellular recordings consists of assigning the recorded spikes to the individual neurons that generated them by grouping them in distinct clusters based on similarities in spike waveform features. High stimulation power interferes with this process by inducing photoelectric artifacts [33, 44, 45, 48] that can distort spike waveforms, especially when short light pulses are used. Additionally, in the case of optogenetic activation (e.g. with channelrhodopsin - ChR2), high light intensities frequently cause synchronous firing in multiple cells, leading to superimposed spike waveforms that cannot be sorted accurately [40, 49].

Another problem is that neurons not expressing opsins may be stimulated indirectly *via* synaptic pathways. This occurs primarily when optogenetic stimulation targets excitatory cells, which can excite post-synaptic neurons upon illumination. This problem is exacerbated with large optical fibers and high light intensities because more neurons are stimulated, increasing the probability that downstream neurons fire. Fortunately, this problem is less serious when ChR2 is expressed in GABAergic inhibitory interneurons, which do not excite postsynaptic cells, although post-stimulus rebound spiking may occur [16, 32]. Other indirect effects such as visual responses evoked by light striking the retina must also be considered [33].

*Proposed improvements for optogenetic identification of interneurons*

Several laboratories have offered solutions for optical stimulation of local circuits combined with simultaneous neuronal recordings. Although these "optrodes" have proved useful in many applications [43-46, 50], at least three improvements can enhance the reliability of optogenetic identification of interneurons: local delivery of low-intensity light, application of appropriate stimulus waveforms, and replacement of large benchtop lasers with small head-mounted LEDs or laser diodes.

Most of the technical problems of unit recording/analysis arise from the use of large-diameter (>100 μm) optical fibers [43-46] or brain surface illumination [13, 47, 48] and the use of high light power. These large fibers are well suited for experiments in which a large illumination area is desired to generate a behavioral phenotype but detrimental for optogenetic circuit analysis and/or extracellular identification of units. Since single-unit spikes can be detected and sorted only ~up to 60 µm laterally from the recording site [49], most of the neurons photostimulated under these conditions are not recorded, making the disambiguation between direct and network-mediated effects more complex. Many of the problems with spike sorting and indirect stimulation can be reduced, if not eliminated, by etching small-core (≤50 µm) optical fibers to a point (~10 µm or less) and mounting them very close (~40 µm) to the recording sites (Fig. 1; [40, 42]). Hybrid devices combining silicon probes or tetrodes with etched optical fibers allow the use of extremely low-power (1-10 µW) stimulation due to the proximity of the recorded neurons to the light source. This method eliminates photoelectric artifacts and enables unequivocal identification of the light-responsive units, while also limiting tissue damage [16, 40, 42].

Overlapping spike waveforms due to synchronous light-evoked action potentials can also be reduced by using structured low-intensity stimulus waveforms such as sinusoids and identifying light-responsive units by their correlation with the stimulus pattern [40, 42]. The rationale behind this approach is that in response to a slowly changing stimulus, targeted neurons do not fire simultaneously because they receive different illumination intensities, express different amounts of opsins, and/or exhibit different resting membrane potentials or thresholds for spike generation. Sinusoidal stimuli have the additional benefit of not inducing photoelectric artifacts that are likely to distort spike waveforms, and they have been used successfully to identify opsin-expressing interneurons *in vivo* [16, 38].

Optogenetics offers unprecedented opportunities to stimulate and silence neurons at multiple locations and structures, which is extremely useful for studying the role of interneurons in ensemble organization [42]. Fine spatiotemporal control of distributed groups of neurons can be achieved by using multiple benchtop lasers coupled through optical fibers to head fixed animals [40, 41, 46]. However, connecting multiple stiff optic patch cords to a small rodent can seriously restrain its movements in tasks that require free navigation. One solution is to use miniature light emitting diodes (LED) and/or laser diodes that are small enough to be mounted on the head of a freely moving animal. These diodes can be coupled to short, small-diameter (≤50 μm) multimode fibers and attached directly to the shanks of a silicon probe or tetrode (Fig. 1, [42]). The small size and weight of these integrated probes (~2 g for a 4-shank/4-LED probe) allow fast, multisite and multicolor optogenetic manipulations in freely moving animals with concurrent monitoring of the manipulated neurons.

The currently method of manually attaching fibers to each probe shank is very labor-intensive and may result in inaccurate alignment. However, efforts are underway for automated fabrication of monolithically integrated optical waveguides and LEDs in multi-electrode silicon probes [51, 52], yielding yet smaller and lighter devices.

**Optogenetics-supervised, physiology-based classification of interneurons**

Over the years, numerous classification schemes based on a variety of physiological criteria were developed to assign extracellular spikes to putative interneurons or pyramidal cells. These include waveform features, firing rate statistics in different brain states, embeddedness in various population activities, firing patterns characterized by their autocorrelograms, and putative monosynaptic connections to other neurons [7, 18-21]. However, it is crucial not only to separate interneurons from pyramidal cells, but also to recognize and correlate activity in different interneuron subtypes with network dynamics and behavior [2, 7, 9]. An important goal of the optogenetic approach is to assist the identification of interneuron classes on the basis of their physiological patterns [7] so that purely physiological criteria could be used in subsequent experiments without the need for optogenetics (Fig. 2) [9]. This would involve iterative refinement of a library of parameters that could be used subsequently for the identification of interneuron subtypes [16, 32, 34, 38]. The optogenetically identified neurons would thus provide the necessary "ground truth" for physiology-based cell type identification. Furthermore, physiological classification methods can also serve to distinguish distinct subtypes of interneurons within individual molecularly identified classes [3].

**Circuit analysis by interneuron perturbation**

Optogenetics not only enables unambiguous identification of interneurons but also provides a way to perturb native network patterns locally and identify the causal role of specific interneuron classes in population activity. Experiments exploiting these perturbation methods are becoming increasingly common in the investigation of interneuron function [4].

For instance, the understanding of visual cortex function has benefited recently from the power of optogenetics. Several studies have addressed the role of specific interneuron subtypes in this region by characterizing their specific response properties and manipulating their activity to determine their impact on principal cell responses. Such optogenetic manipulations suggested that PV interneurons principally control the gain of sensory responses, whereas dendrite-targeting, somatostatin-expressing (SOM) neurons sharpen selectivity [12, 14]. However, other interpretations have also been offered [13, 53]. Notably, SOM interneurons have been shown to play a critical role in surround suppression [26]. Aside from their impact on cortical visual information processing [13, 14, 26, 53], PV interneurons were also shown to be critical for monocular deprivation-induced plasticity [54].

As in the visual cortex, several studies have used optogenetics to decipher the respective functions of two major classes of interneurons, PV and SOM cells, in other brain regions. For instance, large scale extracellular recordings in the prefrontal cortex have shown that PV interneurons exert brief and uniform inhibition on their targets while SOM have longer and more variable inhibitory effects [32]. The same study also demonstrated that a subgroup of optogenetically-identified SOM neurons fired preferentially at reward locations, whereas PV neurons responded when the animal was leaving these reward locations [32]. In the CA1 region of the hippocampus, PV and SOM interneurons exerted complementary effects on place cells by suppressing their activity during the rising and decaying parts of the place field, respectively. Inactivation of PV, but not SOM, interneurons was also shown to interfere with the normal phase assignment of spikes to the theta cycle [34]. Moreover, the strong action of SOM interneurons on spike burst generation in principal cells has been demonstrated both in hippocampus and neocortex, an important feature that is not shared with PV cells [15, 34].

Optogenetics has also advanced the study of interneuron function in oscillatory processes in the neocortex and the hippocampus. In neocortex, for example, strong optogenetic activation of PV-expressing interneurons was shown to amplify gamma oscillations, coordinate the timing of sensory inputs relative to a gamma cycle, and enhance signal transmission [10, 11, 55]. Complementarily, activation of PV interneurons with lower light intensities in both neocortex and hippocampus produced theta resonance and excess spiking in nearby pyramidal cells, demonstrating a specific enhancement of transmission at theta frequency [16]. In the thalamus, repetitive stimulation of the PV neurons of the reticular nucleus switched the thalamocortical firing mode from tonic to bursting, generated state-dependent neocortical spindles [46], and with stronger stimulation evoked generalized spike and wave discharges (Fig. 3) [56]. However, photoactivation of the reticular PV neurons was also found to reduce focal seizures in the neocortex after cortical injury [57]. Similarly, kainic acid-induced seizures could be suppressed by optogenetic activation of PV interneurons in the hippocampus [58]. These recent experiments demonstrate how the power of optogenetics could one day be harnessed for clinical applications, in addition to understanding the role of interneurons in complex cortical functions.

**Outlook**

Optogenetics combined with large-scale extracellular recordings has already proven to be effective in studying the functional roles of specific GABAergic interneuron classes in both hippocampus and neocortex, as well as other brain regions [59, 60]. However, optogenetic identification of interneurons does not allow one to distinguish the different subpopulations of interneurons that belong to a given molecularly defined class (e.g. the subtypes of PV or SOM cells). A further extension of this approach would be to use immediate early gene expression or photoactivatable fluorescent proteins to label light-activated neurons *in vivo* [61]. This labeling would subsequently be used to target these cells for *in vitro* intracellular electrophysiological characterization and/or morphological analysis, providing more detailed information about the cells' identity within each molecular class. Diode probes represent good candidates for this approach because they allow a small number of neurons to be activated selectively and the approximate spatial position of these neurons to be determined based on the silicon probe recording site configuration [62].

Another important extension of current methods is real-time signal processing and closed-loop activation/silencing of interneurons [57, 58]. Illumination could be triggered by spikes of single neurons, combinations of predetermined spike patterns for multiple cells, behavioral parameters, and/or selected features of LFPs [42, 57]. This creates, for instance, the ability to alter timing of action potentials and induce or suppress correlated firing between cells in freely-moving animals. Overall, the combination of optogenetic, large-scale recording and single neuron identification methods will pave the way for a better understanding of the complex dynamics of inhibitory interneurons as well as their roles in coordinating the activity in principal cells in local networks and across network modules.

**Acknowledgements**

We thank S. Royer, A. Berenyi and R. Eichler for their support and E. Schomburg for his comments on the manuscript. Research was funded by National Institute of Health Grants NS34994, MH54671 and NS074015, the Human Frontier Science Program and the J.D. McDonnell Foundation. LR received also support from the Bettencourt Schueller Foundation. LS is supported by NCATS grant UL1 TR000038.

**Figure legends**

**Figure 1. Diode probes for optogenetic identification of interneurons.** A. Schematic of a single LED-fiber assembly. The LED is coupled to a 50-μm multimode fiber, etched to a point at the distal (brain) end. B. Left: schematic of a drive equipped with a 6-shank diode probe with LED-fibers mounted on each shank. Etched optical fibers are attached ~40μm above the recordings sites on the silicon probe shanks. Right: picture of the drive depicted on the left. Scale bar: 3 mm. C-D. Local delivery of light. Magnified frontal view of the 6-shank diode probe equipped with diode-coupled optical fibers. C. Two adjacent shanks illuminated with blue and red light. Scale bar: 1mm. D. Four shanks illuminated with blue light. Scale bar: 0.5 mm (A and C) Reproduced from [42].

**Figure 2. Optogenetic identification of interneurons.** A. Right: unfiltered spontaneous (black) and light-induced (blue) waveforms of a parvalbumin-expressing interneuron (PV) and a pyramidal cell (PYR) at eight recording sites. Note the similarity of the waveforms with and without illumination. Mean and SD; calibration: 0.25 ms, 50 μV. B. Diode probe-induced unit firing in the hippocampal CA1 region (blue shaded area superimposed on the raster plot -top- and the histogram -bottom-; 4 μW at fiber tip). Inset: autocorrelogram shows a shape typical for fast spiking PV interneurons. C. Clustered units are tagged as excitatory or inhibitory based on monosynaptic peaks/troughs in cross-correlation histograms (filled blue and red symbols) and/or response to locally-delivered 50-70 ms light pulses (filled violet symbols) in transgenic mice expressing ChR2 in PV cells. Untagged units (empty symbols) are classified as putative excitatory pyramidal cells (PYR) or inhibitory interneurons (INT) according to waveform morphology; untagged units with low classification confidence are also shown in black ("unclassified") [16]. D. Optogenetic identification of interneuron classes, including here PV- and somatostatin (SOM)-expressing interneurons, allows studying their relationships to network patterns such as sharp wave ripple events. Top: single ripple. Each row represents the color-coded peri-ripple histogram of the firing rate computed for individual neurons. E. Average firing rate observed for the different cell categories. (B) Reproduced from [42]. (C) Reproduced from [16]. (D and E) Reproduced from [34].

**Figure 3. Controlling thalamocortical circuits by optogenetic activation of interneurons.** A. Experimental setup. Optical fiber is placed into the thalamic reticular nucleus in a transgenic mouse expressing ChR2 in PV cells to induce spike-wave seizure-like pattern (shown in C). Blue LEDs (squares) are placed epidurally at two positions in each hemisphere. B. Schematic of the reverberation in the thalamocortical loop. Neurons of the thalamus: reticular nucleus cells (RT), thalamocortical projection neurons (TC). Neurons of the cortex: pyramidal cells (Py) and inhibitory interneurons (Int). D. Light stimulation of the parvalbumin RT neurons alone induces spike-wave discharges, whereas light stimulation of cortical parvalbumin interneurons alone induces rebound excitation in cortical pyramidal cells (Py). Combined and phase shifted stimulation of RT and cortex attenuates the induced spike-wave activity. Reprinted from [56].

**\*\*Juxtacellular recording and labeling of neurons have proven to be a powerful method for to characterize the physiological properties of anatomically characterized interneurons. However, all previous works were carried out in anesthetized animals. This study advance the technique much further by demonstrating that the juxtacellular technique is extendable to behaving animals.**

**\*\* This article demonstrates the power of optogenetics for the identification of two distinct interneuron subtypes in extracellular recordings, to study their respective activity in relation to behavior.**

**\*A cricital requirement for flexible multi-site, multi-color control of interneurons in behaving animals is the small volume and light weight of devices. Integrating waveguides into silicon substrate (see also ref 52) allows for the production of highly flexible recording-stimulation devices.**

62. Buzsaki, G. (2004). Large-scale recording of neuronal ensembles. Nat Neurosci *7*, 446-451.